\documentstyle[twocolumn,prb,aps,epsf]{revtex}
\begin{document}
\def \beq{\begin{equation}}
\def \eeq{\end{equation}}
\def \beqarr{\begin{eqnarray}}
\def \eeqarr{\end{eqnarray}}
\def\etal{{\em et\ al.\/}, }

\twocolumn[\hsize\textwidth\columnwidth\hsize\csname @twocolumnfalse\endcsname
\draft

\title{Sliding Phases via Magnetic Fields}

\author{S. L. Sondhi}
\address{
Department of Physics,
Princeton University,
Princeton, New Jersey 08544
}

\author{Kun Yang}
\address{
National High Magnetic Field Laboratory and Department of Physics,
Florida State University, Tallahassee, Florida 32310
}

\date{\today}

\maketitle

\begin{abstract}

We show that three dimensional ``sliding'' analogs of the Kosterlitz-Thouless
phase, in stacked classical two-dimensional XY models 
and quantum systems of coupled Luttinger Liquids, 
can be enlarged by the application of a parallel magnetic
field, which has the effect of increasing the scaling dimensions of the most
relevant operators that can perturb the critical
sliding phases. Within our renormalization group analysis, we also
find that for the case of coupled Luttinger liquids, this effect is
interleaved with the onset of the integer quantum Hall effect for 
weak interactions and fields. We comment on experimental implications
for a conjectured smectic metal phase in the cuprates.

\end{abstract}
]

\section{introduction}

In 1979, Efetov\cite{efetov} suggested that it would be possible to extend
the low temperature Kosterlitz-Thouless
(KT) phase of a two-dimensional
superconductor to three dimensions by stacking two-dimensional
systems in the presence of a parallel magnetic field. The
underlying idea, most simply understood in a particular gauge
for the field which we specify below, is that the interlayer
Josephson coupling which would ordinarily be relevant even when weak,
is now spatially modulated and no longer gives rise to divergences.
It turns out that this does not work. As pointed out by Korshunov
and Larkin,\cite{larkin} 
the modulated Josephson coupling gives rise to a coupling 
that is {\it not} modulated, and although it is of higher scaling dimension
than the zero field Josephson coupling, it is still relevant
everywhere within the KT phase.
However, recent work on cationic lipid-DNA complexes by O'Hern and
Lubensky \cite{lubensky0} and by Golubovic and Golubovic\cite{gg}, 
and then on XY systems themselves by O'Hern, Lubensky and Toner
(OLT) \cite{lubensky} has found a different way of obtaining analogs
of KT phases in three
dimensions. In this approach, additional derivative couplings
leave the phases in the different planes free to rotate globally
with respect to each other (hence ``sliding phases'') while
extending the region of irrelevant vortex fugacity to a range
where the interlayer Josephson coupling is now irrelevant.
Emery \etal \cite{emery} and Vishwanath and Carpentier \cite{vc}
have applied this insight to quantum problems and obtained 
an analog of the Luttinger liquid in two dimensions.

Our purpose in this note is to point out that one can combine
Efetov's insight with the more recent work and considerably
extend the domain of these sliding phases by reducing the 
dimension of a large class of relevant operators via the action 
of a parallel magnetic field. This is of considerable interest
for the full class of perturbations in such problems can be
quite constraining,\cite{vc} even though it is reasonable that most
of them are not realized with substantial amplitude \cite{emery}.
We will be especially interested in ``sliding Luttinger
liquids'' or ``smectic metals'' which have been argued to arise
in the cuprate superconductors on account of the stripe instability
of a doped antiferromagnet \cite{emery,fekiv}. \
We should note that there is a close
connection between our work and that on the striped phases in
high Landau levels \cite{qhstripes} even though our 
point of departure (Efetov's conjecture) is very different. 
In the Landau level problem, the
field is built in at the first step and is central in giving
rise to the striped phase in the first instance, while for us
it can be variable in magnitude and give rise to both
gapped quantum Hall and gapless smectic behavior and an
interesting phase transition between them. Nevertheless,
in both cases the field serves to constrain the available set
of relevant operators in very similar fashion. 

We will begin in Section II with a quick account of the 
``dimensional reduction'' of the Josephson coupling produced by a 
parallel field, the genesis of the sliding phase and its enlargement 
by the field. Next (Section III) we discuss the application of these 
ideas to coupled Luttinger liquids and present contrasting phase diagrams
for a model studied by Emery {\em et al.} In this discussion we also
show how the integer quantum Hall states are rediscovered by
perturbation theory about a smectic metal if the interactions
are not too strong. We close with a brief summary and a discussion
of possible experimental implications for the cuprates.

\section{Sliding XY Phases in Parallel Fields}

We begin with a brief summary of the genesis of the sliding phase
in a three dimensional stack of layers characterized by and XY
order parameter. We largely follow OLT 
and their notation for ease of comparison. The Hamiltonians of
the sliding phase fixed points (the plural is warranted) belong to
the family,
\begin{equation}
\label{sliding}
H_S  = {1 \over 2} \sum_{n n'}\int d^2 r~K_{n n'}
{\mbox{\boldmath{$\nabla$}}}_{\perp} \theta_n({\bf r}) \cdot
{\mbox{\boldmath{$\nabla$}}}_{\perp} \theta_{n'}({\bf r}), \\
\end{equation}
where $K_{n n'} = K f_{n - n'}$ with $f_n = (1 + \sum_m \gamma_m)
\delta_{n,0} - \case{1}{2} \sum_m \gamma_m (\delta_{n,m} + \delta_{n,-m})$
and and ${\boldmath \nabla}_{\perp} \theta_n({\bf r})$ 
denotes the in-layer gradient of the XY variable in layer $n$.
We take ${\bf r} \equiv (x,y)$ and set the separation of successive layers
along the $z$-axis to 1.
One can check that $H_S$ is invariant under 
shifts $\theta_n({\bf r}) \rightarrow \theta_n({\bf r}) + \psi_n$ for any 
choice of $\psi_n$. This freedom to globally rotate the angle in one
layer relative to another, even in the presence of interlayer couplings
in $H_S$, is the hallmark of the sliding phase.

Note that $H_S$ returns to itself under a renormalization group (RG)
transformation that ``lives'' in two dimensions and treats the
layer index $n$ as an internal or flavor index on the fields
$\theta_n$. In order to identify functions $K_{n n'}$ that would
govern {\it stable} fixed points under this RG, we need to examine
the behavior of vortex fugacities and Josephson couplings. The
former yield, for a vortex configuration $\{\sigma_n\}$ in which
a net vorticity $\sigma_n$ occurs in layer $n$, the scaling dimension
\begin{equation}
\Delta_{v}[\sigma_n] ={\pi K \over T} \sum_{n,n'} f_{n-n'} \sigma_n
\sigma_{n'} \ ,
\label{etasig}
\end{equation}
which signals a KT unbinding transition at a
temperature $T_{KT}[\sigma_n]$, upon exceeding the value 2 
appropriate to a two-dimensional RG. The generalized
Josephson couplings,
\begin{equation}
\label{Josephson}
H_J[s_n] = - V_J[s_n] \int d^2r \cos\left[
\sum_p s_p \theta_{n+p}
({\bf r})\right],
\end{equation}
where the $s_n$ are integers that satisfy $\sum_n s_n =0$, are readily
shown to have the scaling dimension,
\beq
 \Delta_J[s_n] =
{T \over 4 \pi K} \sum_{n,n'} s_n s_{n'} f^{-1}_{n - n'} \ ;
\eeq
where the inverse couplings
\begin{equation}
f_p^{-1} = {1 \over \pi} \int_0^{\pi} dk {\cos kp \over f(k)} .
\label{fp}
\end{equation}
are defined via the Fourier transform
\begin{equation}
f(k) = 1 + \sum_m \gamma_m (1 - \cos km )
\end{equation}
of the scaled couplings $f_n$.

The Josephson couplings are irrelevant above a decoupling temperature 
$T_d[s_n]$ at
which $\Delta_J[s_n] = 2$. If $\min_{\sigma_n} T_{KT}[\sigma_n]
> \max_{s_n} T_d[s_n]$ for some choice of $K_{n n'}$ then we obtain
a sliding phase.
In the sliding phase the spin correlations are algebraically long
ranged in a given layer and vanish between layers,
\beq
\langle \cos[\theta_n ( {\bf r} ) - \theta_m ( 0)] \rangle
\sim {\delta_{nm} \over r^\eta} \ ,
\eeq
where $\eta = {T \over 2\pi K} f_0^{-1}$.

\noindent
{\bf Including a parallel magnetic field:} We now consider the
inclusion of a magnetic field parallel to the layers,
appropriate to instances where the $\theta_n$ are 
phases of a superconducting order parameter; without loss of
generality, we take ${\bf B} = B \hat{y}$. It is convenient to 
work in the gauge $A_z ({\bf r}) = B x$. In this gauge,
the sliding phase Hamiltonians are of the same form, and the
computation of the scaling of the vortex fugacity is unchanged.
However, the Josephson couplings are modified by the replacements
\beq
 \theta_n \rightarrow \theta_n + 2n q_B x
\eeq
where  $q_B = { e B  \over \hbar c}$ is a characteristic wavevector 
introduced by the field. 

The key observation regarding the effect of the field is this:
for those Josephson couplings for which $\sum_p p s_p \ne 0$,
there is an explicit oscillating term in the argument of the
cosine that will render them less relevant. Most straightforwardly,
consider treating such a term in perturbation theory. In zero
field, we would discover that the term was relevant upon finding
divergences in perturbation theory. The inclusion of the field
will attenuate these divergences due to the oscillation of the
correlation functions of the perturbation.

However, the net result is not always to render the
perturbation theory convergent. Higher order graphs can
involve regions where products of the oscillating couplings
nevertheless give rise to operators that do not oscillate.
For example, the Josephson coupling for layers at distance
$p$,
\beqarr
\cos[ \theta_{n+p}({\bf r}) - \theta_{n}({\bf r}) &+& 2p q_B x]
\nonumber \\
&\sim& e^{i[\theta_{n+p}({\bf r}) - \theta_{n}({\bf r})]}
e^{i2p q_B x} + {\rm c.c.} 
\label{twolayerterm}
\eeqarr
will give rise to
\beq
\cos[ \theta_{n+p}({\bf r}) - 2 \theta_{n}({\bf r})
+ \theta_{n-p}({\bf r})]
\label{threelayerterm}
\eeq
which can then produce divergences of its own. Indeed, this
particular generation is exactly what invalidates Efetov's
original conjecture, for the operators (\ref{threelayerterm}) 
are relevant
everywhere in the KT phase of decoupled XY layers. Nevertheless,
the application of the field does effect a ``dimensional
reduction'' in that ``charged'' operators that have a net 
$\sum_p p s_p$ (microscopically these arise from hopping 
processes that move a net charge up or down the stack) can only 
affect the result through the generation of ``neutral'' operators 
for which  $\sum_p p s_p=0$. At the (unstable) decoupled 2D XY 
fixed line, the latter have higher dimension and we might expect 
that this will be true at sliding fixed points as well. While that 
is not always the case, as will be clear by the following example,
it is still
the case that knocking out the charged operators improves the
stability of the sliding phase---after all, the neutral operators
were present anyway!

\begin{figure*}[h]
\centerline{\epsfxsize=9cm
\epsfbox{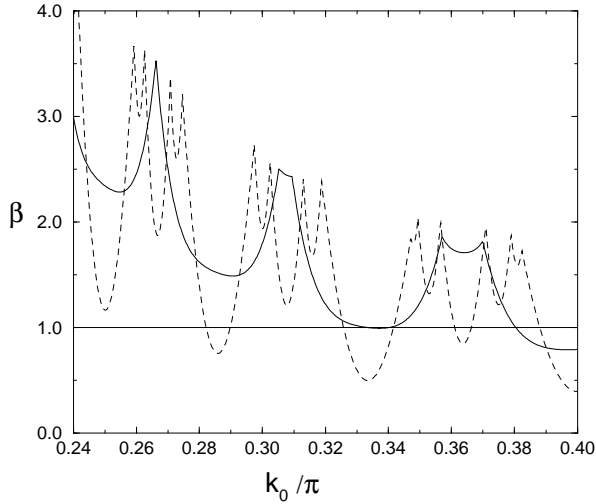}
}
\caption{
A plot of $\beta=T_{KT}/T_d$ against $k_o/\pi$ at $\delta = 10^{-5}$.
The dashed line
is due to charged two-layer couplings while the solid line is
due to the leading, three layer, neutral couplings. The $B=0$
sliding phase exists when the minimum of the two curves exceeds
1. The $B \ne 0$ sliding phase requires only that the solid
curve exceed unity and hence leads to a larger sliding phase.
}
\label{fig:xy}
\end{figure*}

To illustrate this effect, we consider the example used by OLT
with first and second neighbor couplings. The coupling function
$f_n$ has a Fourier transform
\begin{equation}
\label{Kqz}
f(k) = 1 + \gamma_1 (1 - \cos k ) + \gamma_2 (1 - \cos 2k )
\end{equation}
that is required to take its minimum value at $k = k_o$:
\beq
f(k_o)= \delta, \ \ \ f'(k_o)=0 \ {\rm and}\  f''(k_o)= 2 C  \ .
\eeq
Sliding phases arise when $\delta$ and $k_o$ are chosen so that
the system is close to an incommensurate transverse ordering 
instability, as has been discussed nicely by Vishwanath and 
Carpentier \cite{vc}. At small $\delta$, the asymptotic form,
\beq
 f^{-1}_p \sim {\cos(pk_0)e^{-p \sqrt{\delta/C}} \over 
\sqrt{C \delta}}
\eeq
enables easy numerical calculation of the scaling dimensions
of the Josephson couplings and thence of the temperatures $T_d$. 
In Fig. \ref{fig:xy}, 
we plot the ratio, $\beta= \min_{\sigma_n} T_{KT}[\sigma_n]
/\max_{s_n} T_d$ where the $s_n$ are restricted to the two layer 
Josephson couplings (\ref{twolayerterm}) and the three layer terms
that they generate (\ref{threelayerterm}). 
The value of $k_o$ where {\it both} are greater than 1 support a sliding phase
in zero field \cite{fn-olh}, while the latter
alone determines the sliding phase in a magnetic field. The
expansion of the phase is clear. (We have not attempted to include all 
operators that might be allowed by symmetry. As noted in 
Ref.~\onlinecite{vc} in 
the Luttinger liquid context, higher order operators allow increasingly
finer instabilities. We do not know of a proof that all such operators
allow or exclude a connected sliding phase, but assume that in
a given system a finite set will be important over some reasonable
range of length scales. Regardless, the magnetic field will improve
matters by knocking out all the charged operators.)

\section{Smectic Metals in Transverse Fields}

In this section we discuss coupled one-dimensional (1D) Luttinger liquids 
(LL) in the presence of
a magnetic field, with the field ${\bf B}$ transverse to the plane 
in which the 1D chains are placed. It was known that at the {\em decoupled} LL 
fixed points, 
the transverse inter-chain coupling is always relevant, in one of 
three channels: single electron hopping, (Cooper) pair hopping, and inter
chain $2k_F$
back scattering. As a consequence, the decoupled LL phase is always unstable 
and driven toward the Fermi liquid \cite{fn-pwa}, superconducting, or 
charge/spin density wave (CDW/SDW) phases. It was recently pointed 
out\cite{emery,vc} that 
adding strong interchain forward scattering terms (which are exactly
marginal) to the decoupled LL fixed point can drive all these interchain 
couplings irrelevant. The resulting stable, non-Fermi liquid, smectic metal 
phase is the quantum analog of the classical sliding phase.\cite{lubensky}

Since the single electron and Cooper pair hopping processes involve charge
transfer between neighboring chains, the presence of a magnetic field has a
similar effect, as before, of increasing the scaling dimensions of the 
operators corresponding to these processes
that can perturb the smectic metal fixed points, and hence increasing the 
range of stability of the smectic metal phase (for simplicity we will 
neglect the Zeeman effect of the field in this paper). 
In the following we present an explicit
analysis of this effect. Following Emery \etal \cite{emery} three different 
types of smectic metal fixed points
need to be distinguished and analyzed in turn: \\
(i) a spinful smectic metal with a spin gap; \\
(ii) a spinful smectic metal without a spin gap; and \\
(iii) a spinless smectic metal. \\
For simplicity we will only include nearest neighbor interchain
couplings and their immediate descendants. 

\begin{figure*}[h]
\centerline{\epsfxsize=9cm
\epsfbox{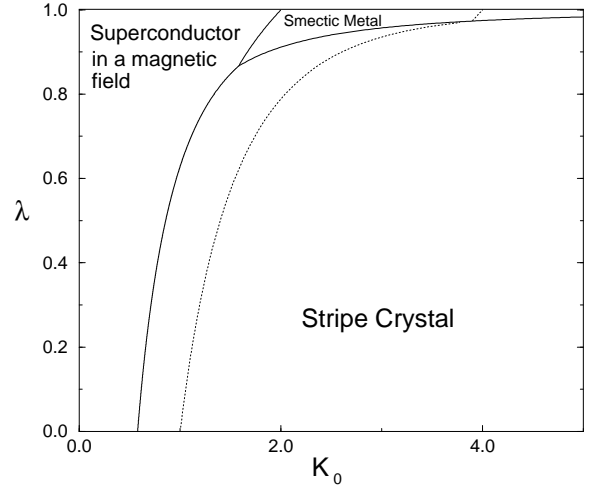}
}
\caption{
Phase diagram for coupled Luttinger liquids with a spin gap, in the presence of
a magnetic field. For comparison we have also included the phase boundaries 
separating the superconducting phase from the smectic metal and the
stripe crystal
phases in the {\em absence} of a magnetic field (dotted lines), 
{\em a la} Emery {\em et al.} \cite{emery} The presence of a magnetic
field significantly expands the region of both the smectic metal and the
stripe crystal.
} 
\label{fig:gap}
\end{figure*}

\noindent
{\bf Spin-gapped smectic metal}. 
In this case the fixed point action in Euclidean space takes
the form\cite{emery}
\beqarr
S&=&{1\over 2}\sum_Q[W_0(k_\perp)\omega^2+W_1(k_\perp)k^2]
|\phi(Q)|^2\nonumber\\
&=&{1\over 2}\sum_Q[{\omega^2\over W_0(k_\perp)}+{k^2\over W_1(k_\perp)}]
|\theta(Q)|^2,
\label{sgaction}
\eeqarr
where $Q=(\omega, k, k_\perp)$, for each chain the 2-current is 
$j_\mu={1\over \sqrt{\pi}}\epsilon_{\mu\nu}\partial^\nu \phi$, and $\theta$ is
the dual field of $\phi$. The scaling dimensions of various local operators 
are determined by the dimensionless Luttinger coupling function
\beq
w(k_\perp)=\sqrt{W_0(k_\perp)W_1(k_\perp)},
\eeq
which is periodic in $(k_\perp)$ with period $2\pi$ as we have set the
interchain distance to 1.
As in Ref.\onlinecite{emery}, we consider the simplified model in which
$w(k_\perp)$ takes the form
\beq
w(k_\perp)=K_0+K_1\cos(k_\perp)=K_0[1+\lambda\cos(k_\perp)].
\eeq
Stability requires $|\lambda| < 1$.
In the presence of a spin gap single electron hopping is irrelevant, and
the magnetic field has no effect on $2k_F$ back scattering which does not
involve charge transfer between chains. We thus focus on the singlet pair
hopping process, which in the presence of a magnetic field is described by the
following perturbing Hamiltonian (near neighbor hopping only):
\beqarr
H_{sc}&=&-t_J\int{dx}\, h_{sc}(x),\nonumber\\
h_{sc}(x)&=&\sum_j\cos[\sqrt{2\pi}(\theta_j(x)-\theta_{j+1}(x))+2q_Bx],
\eeqarr
where $t_J$ is the Josephson coupling strength and $q_B=eB/\hbar c$ as
before. As in the previous section, the field adds an oscillatory phase 
to the pair hopping term, which renders $h_{sc}$ irrelevant by itself.
However, as it flows, it again generates terms in which the oscillatory 
phases cancel. The most relevant of these is 
\beqarr
\tilde{H}_{sc}&\propto& t^2_J\int{dx} \tilde{h}_{sc}(x),\nonumber\\
\tilde{h}_{sc}(x)&=&\sum_j\cos[\sqrt{2\pi}(2\theta_j(x)
-\theta_{j+1}(x)-\theta_{j-1}(x))],
\label{secondscterm}
\eeqarr
which is generated at second order in $t_J$. The scaling dimension of this
term is
\beqarr
\tilde{\Delta}_{sc}&=&{2K_0\over 2\pi}\int_0^{2\pi}{dk_\perp}
(1+\lambda\cos(k_\perp))(1-\cos(k_\perp))^2\nonumber\\
&=&(3-2\lambda)K_0.
\eeqarr
Combining the knowledge of $\tilde{\Delta}_{sc}$ with the scaling dimension of
the $2k_F$ back scattering operator\cite{emery}
\beq
\Delta_{CDW}={2\over K_0(1-\lambda+\sqrt{1-\lambda^2})},
\eeq
we can determine the phase diagram of the model (\ref{sgaction}) in the 
presence of a magnetic field and near neighbor interchain couplings,
and subject to weak, generic perturbations, 
using the criteria that the smectic phase is stable when $\tilde{\Delta}_{sc}
> 2$ and $\Delta_{CDW}> 2$; otherwise the system is in the 
stripe crystal/superconducting phase for $\Delta_{CDW}$ smaller/bigger than
$\tilde{\Delta}_{sc}$. (In this identification we have made the natural
assumption that the coupling (\ref{secondscterm}) will govern the 
properties of the phase when it grows most rapidly. By itself, it will
produce a vortex lattice \cite{radbal}.)
The phase diagram is
plotted in Fig.~\ref{fig:gap}. For comparison we have also included 
the phase boundaries
separating the superconducting phase from the smectic metal and stripe crystal
phases in the {\em absence} of a magnetic field\cite{emery} as dotted lines.
It is quite obvious that both the smectic metal and stripe
crystal phases get expanded by the magnetic field, which suppresses interchain
Josephson coupling and increases the scaling dimension of operators involving
pair hopping.

\noindent
{\bf Spin-ungapped smectic metal}. In this case the fixed point action has 
contributions from both the charge and spin sectors: $S=S_\rho+S_\sigma$,
where we take $S_\rho$ to have the same form as Eq. (\ref{sgaction}), and 
\beq
S_\sigma={K_\sigma\over 2}\sum_j[{1\over v}(\partial\tau\phi_{j\sigma})^2+
v(\partial_x\phi_{j\sigma})^2],
\eeq
in which we assume that there is no interchain coupling among spin fields, 
as in Ref. \onlinecite{emery}. Spin rotation invariance (also assumed here)
requires $K_\sigma=1$. The analysis of pair hopping is similar to the
previous case and it is easy to show that the most relevant 
operator generated by the pair hopping has scaling dimension 
$\tilde{\Delta}_{sc}^{nogap} =\tilde{\Delta}_{sc}^{gap}+3 > 2$; {\em i.e.}, 
operators generated by pair hopping are {\em always irrelevant} here.

Low-energy single electron hopping, which is allowed, is on the other hand
more complicated and interesting. In terms of the original electron operators
it takes the form
\beqarr
H_e&=&-t_e\int{dx}\, h_e(x),\nonumber\\
h_e(x)&=&\sum_{j\sigma}(\psi_{j\sigma}^\dagger(x)\psi_{j+1\sigma}(x)e^{iq_Bx}
+h.c.).
\eeqarr
We need to distinguish two different cases here. 

\begin{figure*}[h]
\centerline{\epsfxsize=9cm
\epsfbox{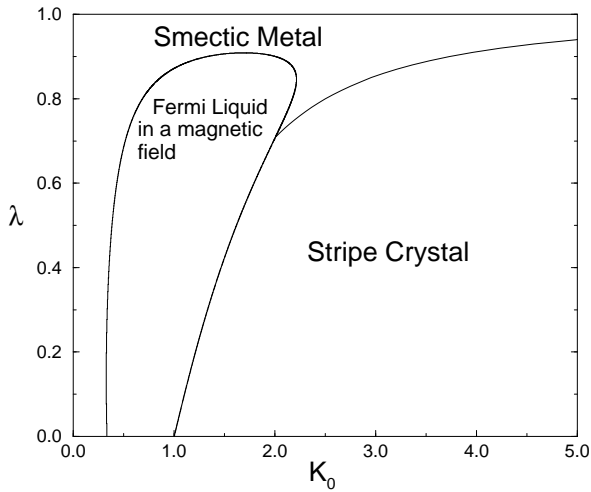}
}
\caption{
Phase diagram for coupled spinful Luttinger liquids without a spin gap, 
in the presence of
a magnetic field, assuming there is no commensuration between $k_F$ and $q_B$.
}
\label{fig:nogap}
\end{figure*}

(i) $k_F$ and $q_B$ are incommensurate.
In this case single-electron processes all involve an oscillating phase, and 
the most relevant process without an oscillating phase generated by $H_e$ is
\beqarr
\tilde{H}_e&\propto& t_e^2\int{dx}\,\tilde{h}_e(x),\nonumber\\
\tilde{h}_e(x)&=&\sum_i[\psi_{j\uparrow}^{L\dagger}(x)
\psi_{j\downarrow}^{R\dagger}(x)\psi^L_{j+1\uparrow}\psi^{R}_{j-1\downarrow}
+h.c.+\cdots],
\eeqarr
where $L/R$ stands for left/right mover, and $\cdots$ stands for terms of 
similar structure. In bosonized form,
\beqarr
& & \tilde{h}_e(x)\propto \\
& &  \cos\{\sqrt{2\pi}[2(\theta_{\rho i}+\phi_{\sigma i})
-\theta_{\rho i+1}-\theta_{\rho i-1}-\phi_{\sigma i+1}-\phi_{\sigma i-1}]\},
\nonumber 
\eeqarr
which has the scaling dimension
\beq
\tilde{\Delta}_e=1+{K_0\over 2}({3\over 2}-\lambda)+{1-\sqrt{1-\lambda^2}
\over 2K_0\lambda^2\sqrt{1-\lambda^2}}.
\eeq
We assume that the system behaves as a Fermi liquid in a magnetic field
when this term dominates. This identification is suggested if we note 
that at the non-interacting point, $\lambda=0$ and $K_0=1$, this
term is marginal.
This leads to the phase diagram Fig. \ref{fig:nogap}, which is qualitatively
different from the phase diagram in the absence of the field, Fig.~2 of 
Ref.~\onlinecite{emery}. There are two particularly interesting differences:
i) The superconducting phase gets completely squeezed out by the field;
ii) The smectic metal phase now extends all the way to $\lambda=0$, which 
corresponds to the {\em decoupled} LL fixed point, a situation impossible 
without the field.

\begin{figure*}[h]
\centerline{\epsfxsize=9cm
\epsfbox{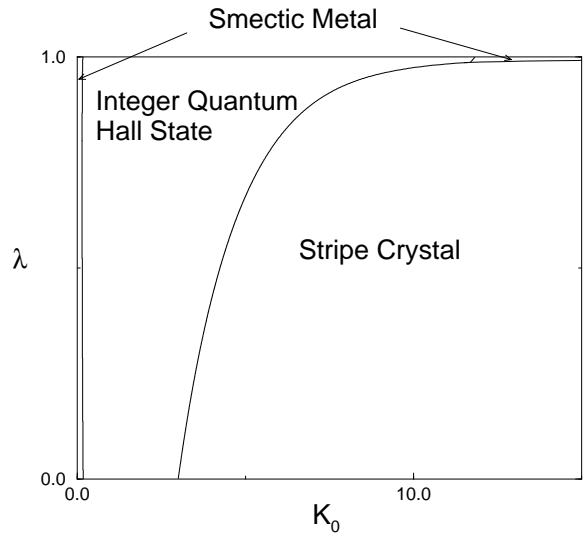}
}
\caption{
Phase diagram for coupled spinful Luttinger liquids without a spin gap,
in the presence of a magnetic field, in the commensurate case
$2k_F=nq_B$.
}
\label{fig:nogapqh}
\end{figure*}

(ii) $2k_F=nq_B$ where $n$ is an integer.
In this case $H_e$, or its higher order descendents in the low energy theory,
can turn a left mover on the Fermi point of the $j$th chain to a right mover
on the Fermi point of the $j+n$th chain; this is a low energy single electron
hopping process that {\em does not} involve an oscillatory phase, which takes
the form
\beqarr
H'_e&=&-t_e\int{dx}h'(x),\nonumber\\
h'(x)&=&\sum_{j\sigma}(\psi^\dagger_{jL}\psi_{j+nR}+h.c.)\nonumber\\
&\propto& \cos\sqrt{\pi\over 2}(\theta_{\rho i}-\theta_{\rho i+n}
+\phi_{\rho i}+\phi_{\rho i+n})\nonumber\\
&\times&\cos\sqrt{\pi\over 2}(\theta_{\sigma i}-\theta_{\sigma i+n}
+\phi_{\sigma i}+\phi_{\sigma i+n}).
\eeqarr
The scaling dimension of this operator $\Delta'_{e,n}$ for $n=1$ is 
\beq
\Delta'_{e,1}={K_0\over 4}(1-{\lambda\over 2})+{1\over 2K_0(1+\lambda
+\sqrt{1-\lambda^2})}+{1\over 2}.
\eeq
In regions of parameter space where this is the most relevant operator,
we expect that the system develops a gap that is largely single particle
in character. The identification of the resulting state is easy once we
recognize that the condition $2k_F=nq_B$ is precisely that the Landau level
filling of the system is $\nu=2n$---i.e. the electrons (inclusive of
their spin degeneracy) occupy $n$ Landau bands and form an integer
quantum Hall state!

In Fig.~\ref{fig:nogapqh}
we show the phase diagram for the case of $n=1$ ($\nu=2$). As the transition 
between the quantum Hall state and the smectic metal happens via
the hopping going irrelevant, it is a continuous transition. To
our knowledge, this is the first instance of a continuous transition
between a quantum Hall state and a metallic state. We should note that
the persistence of the quantum Hall phase upto the upper boundary at
$\lambda=1$ is non-generic; it arises in the particular model studied
upon a cancellation between numerator and denominator that will not
typically take place.

Finally, higher order commensurations between $k_F$ and $q_B$ are possible
when lattice effects are strong on the chains and the electron operator
has pieces oscillating at higher multiples of $k_F$. We have not
investigated these.

\begin{figure*}[h]
\centerline{\epsfxsize=9cm
\epsfbox{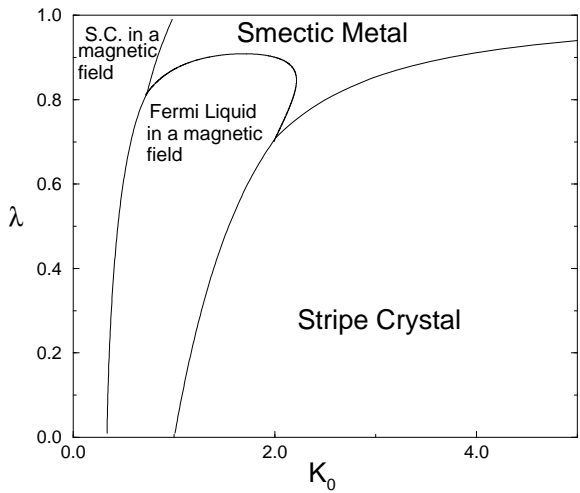}
}
\caption{
Phase diagram for coupled spinless Luttinger liquids, in the presence of
a magnetic field, assuming there is no commensuration between $k_F$ and $q_B$. 
}
\label{fig:spinless}
\end{figure*}

\noindent
{\bf Spinless smectic metal}. In this case we only have charged fields as
in the spin gapped case, but single electron processes need to be considered
as in the spin ungapped case. The analysis of perturbing operators is very 
similar to the spin ungapped case, which leads to the phase diagram Fig.
\ref{fig:spinless} when $k_F$ and $q_B$ are incommensurate. Integer quantum
Hall cases are, of course, allowed here as well, when $2k_F=nq_B$.

\noindent
{\bf Disorder:} Following Giamarchi and Schulz \cite{gs}, one can also
analyze the scaling of weak single-particle randomness. We have not done
this systematically, but will content ourselves with a couple of remarks.
First, in all cases it is possible to find subsets of the smectic metal
where both intrachain random backscattering and interchain random hopping
are {\em irrelevant}---hence the system is a perfect, albeit completely
anisotropic,
metal in the long-wavelength limit. Second, it is possible to find
sections of the phase boundary between the quantum Hall states and the
smectic metal where disorder is still irrelevant, e.g. in the spin
ungapped problem this happens both near $\lambda=0$ and near $\lambda=1$.
In these cases we find an analytically tractable fixed point governing
a transition out of a quantum Hall state in the presence of interactions
{\it and} disorder that warrants further analysis \cite{fn-ye}. 
 
\section{Summary}

Achieving a ``dimensional continuation'' of strong
correlation physics from low dimensions by weakly coupling 
an infinite set of systems is an appealing strategy 
in the study of higher dimensional systems \cite{dimcont}.
The application of a magnetic field has been conjectured
previously to be useful in this task. In addition to the
work of Efetov, we should also mention the suggestion of
Strong, Clarke and Anderson \cite{sca} that a two-dimensional
non-Fermi liquid phase could be induced in this fashion
in a layered system. Striking experiments in the organic
superconductors that are evidence for this point of view
have been discussed at some length \cite{sc}.

In this paper, we have shown that this decoupling effect
of the magnetic field can be given precise meaning in
the context of two-dimensional sliding phases, via its
reduction of the dimension of the most relevant charged
operators that perturb them. This significantly expands
the size of the sliding phases. As a bonus we find,
in the quantum version of the problem, quantum Hall phases 
at commensurate fields that undergo a novel continuous 
transition to a smectic metal.

In the underdoped region of the cuprates, it has been
argued that the stripe instability leads to a smectic metal 
state and that it may already have been observed\cite{emery}. 
In this setting, the spin gapped phase discussed here is
the one at issue, whence we anticipate that the Zeeman
coupling (ignored in our analysis) will not be important. 
We suggest that the field sensitivity of the phase diagram
in this region would be an interesting test of the 
smectic hypothesis---essentially, one should look for
the expansion of the metal or the onset of a CDW. The
parameters needed to see this effect should ensure that
the interchain hopping is weaker than the field, 
$t_e < v_F q_B$ ($v_F$ is the on-chain Fermi velocity)
and that the temperature does not wash out
the phases induced by the field. The latter condition
can be translated, via the on-chain smectic correlation 
length $\xi \sim {\epsilon_F/n T}$  ($\epsilon_F$ is the
on chain Fermi energy and $n$ is the linear density
of electrons), to the statement $B a \xi \sim \phi_o$
where $a$ is the interchain spacing and $\phi_o$
is the flux quantum.

\acknowledgements
We are grateful to Ashvin Vishwanath, Tom Lubensky and Steve Kivelson 
for useful discussions. 
We are especially grateful to David Huse for many illuminating
discussions and especially for producing a lucid, intuitive
explanation of the content of Ref.~\onlinecite{larkin}.
This work was supported by NSF DMR-9971541 and the Sloan Foundation
(KY), and NSF DMR-9978074 and the Sloan and Packard Foundations (SLS).


\begin{references}
\bibitem{fn-olh} Readers wishing to compare our Fig.~\ref{fig:xy} with 
Fig.~1 in Ref.~\onlinecite{lubensky} should note that the latter does not 
include the instabilities
due to the three layer operators in Eqn.~(\ref{threelayerterm}). 
The statement in Ref.~\onlinecite{lubensky} that the latter are always 
less relevant is not correct.
\end{references}
\end{document}